\documentclass{PoS}

\title{Is the generalized Brink-Axel hypothesis valid?}

\ShortTitle{The Brink-Axel hypothesis}

\author{\speaker{M. Guttormsen}$^a$, A. C. Larsen$^a$, A.~G{\"o}rgen$^a$,
T.~Renstr{\o}m$^a$, S. Siem$^a$, T. G. Tornyi$^{a,b}$ and G. M. Tveten$^a$\\
\llap{$^a$}Department of Physics, University of Oslo, Norway\\
\llap{$^b$}Department of Nuclear Physics, Australian National University, Canberra, Australia\\
        E-mail: \email{magne.guttormsen@fys.uio.no}}

\abstract{Experimental results of the $^{237}$Np($d, p \gamma)^{238}$Np reaction are presented,
which verifies the generalized Brink-Axel (gBA) hypothesis for $\gamma$ transitions between
states in the quasi-continuum. The gBA hypothesis holds not only for specific
collective resonances, but for the full dipole strength below the neutron
separation energy. We discuss the validity of the gBA hypothesis also for lighter
systems like $^{92}$Zr where the concept of a unique $\gamma$-ray strength function ($\gamma$SF)
is problematic due
to large Porter-Thomas fluctuations. Methods for studying the $\gamma$SF
and the fluctuations as function of excitation energy are presented.}

\FullConference{The 26th International Nuclear Physics Conference\\
		11-16 September, 2016\\
		Adelaide, Australia}

\begin{document}

\section{Introduction}
More than sixty years ago, Brink proposed in his doctoral
thesis at Oxford University~\cite{brink,oslo2009}
that the photoabsorption cross section of the giant electric dipole
resonance (GDR) is independent of the detailed structure of the initial
state. The hypothesis was further extended by the principle of detailed balance,
to include absorption and emission of $\gamma$ rays between resonant states~\cite{axel,Bartholomew}.
In more general terms, this generalized Brink-Axel (gBA) hypothesis implies that the
dipole $\gamma$-decay strength has no explicit dependence on excitation
energy, spin or parity, except the obvious dipole transition selection rules.

The gBA hypothesis is frequently applied in a variety
of applications as it dramatically simplifies the considered problem
and in some cases, is a necessity in order to perform the calculations.
Hence, the question of whether the hypothesis is valid or not, and
under which circumstances, is of utmost importance.
For this discussion, one should keep
in mind that the original formulation of the hypothesis was meant for
moderate energies, as cited from Brink's thesis: {\em "Now we assume that the
energy dependence of the photo effect is independent of the
detailed structure of the initial state so that, if it were possible
to perform the photo effect on an excited state, the cross section for
absorption of a photon of energy $E$ would still have an energy dependence
given by (15)"}, where the equation (15) refers
to a Lorentzian shape of the photoabsorption cross section on the ground state.

The gBA hypothesis has fundamental impact on nuclear structure and dynamics,
and has a pivotal role in the description of $\gamma$ and $\beta$ decay for
applied nuclear physics. In particular, the hypothesis is often used
for calculating ($n, \gamma$) cross-section needed to model
the astrophysical $r$-process nucleosynthesis and the next-generation
of fast nuclear reactors.

Experimental and theoretical information on the validity of the gBA hypothesis is rather scarce.
Primary transitions from $(n,\gamma)$ reactions give
$\gamma$SFs consistent with the gBA hypothesis, but only in a limited
$\gamma$-energy range~\cite{stefanon1977,raman1981,kahane1984,islam1991,kopecky1990}.
On the other hand,  the $^{89}$Y$(p,\gamma)^{90}$Zr
reaction points towards deviations from the gBA hypothesis~\cite{netterdon2015}.
There have also been theoretical attempts to test the gBA hypothesis. Here, deviations and
even violations to the hypothesis are found~\cite{johnson2015,misch2014}.
However, for other theoretical applications, the assumption
of the gBA hypothesis is successfully applied~\cite{koeling1978,horing1992,gu2001,betak2001,hussein2004}.

From the above mentioned findings, it is not at all obvious when the gBA hypothesis is valid.
This unclear situation is not only due to peculiar structures and dynamics of the systems studied,
but also in some cases, due to huge Porter-Thomas fluctuations~\cite{PT} that may camouflage the underlying
physics. In this work, we will show that at moderate excitation energies and with a
proper averaging over many $\gamma$ transitions, the gBA hypothesis is a fruitful
concept for the exploration of the nuclear quasi-continuum region.

\begin{figure}[]
\begin{center}
\includegraphics[clip,width=\columnwidth]{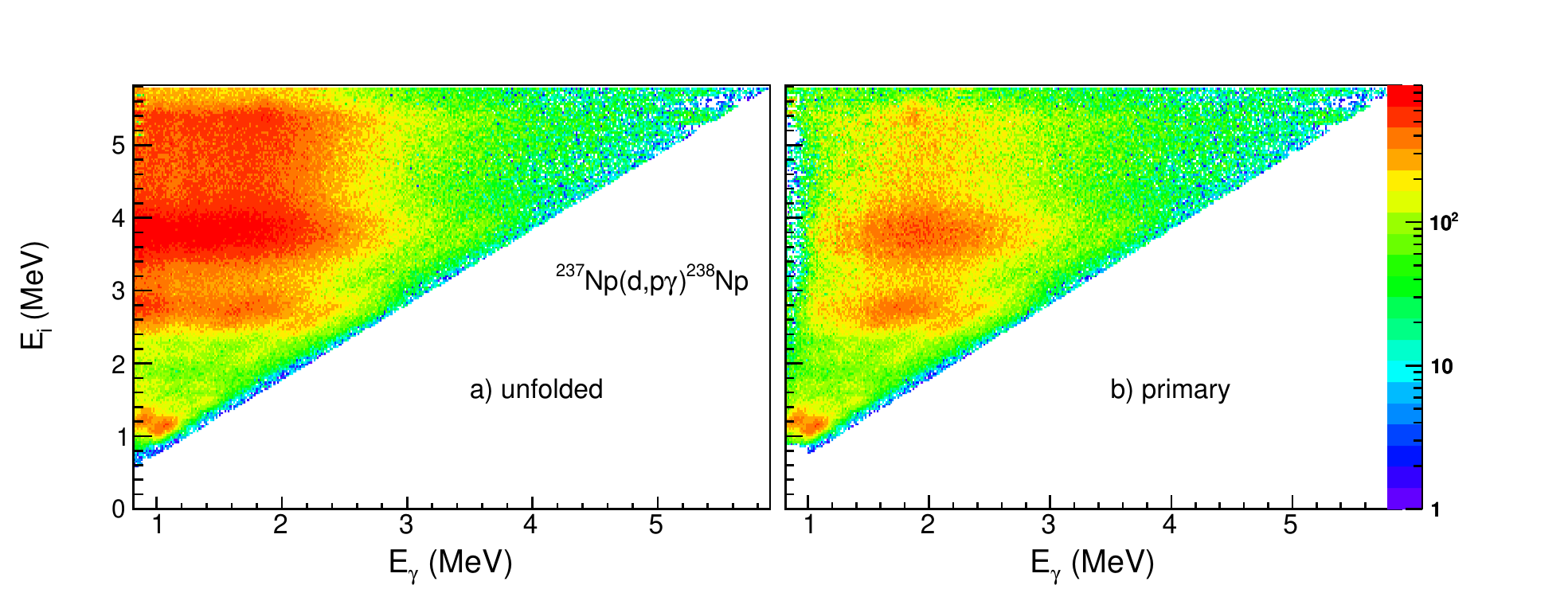}
\caption{(Color online) Initial excitation energy $E_i$ versus $\gamma$-ray energy
$E_{\gamma}$ from particle-$\gamma$ coincidences recorded with the
$^{237}$Np$(d,p\gamma)^{238}$Np reaction~\cite{238Np,guttormsen2016}. The  $\gamma$-ray spectra
unfolded by the NaI response function (a) and the primary or first-generation
$\gamma$-ray spectra $P(E_\gamma,E_i)$ (b) are extracted as function of initial excitation energy $E_i$.}
\label{fig:matrix}
\end{center}
\end{figure}

\section{Method and tools}

The Oslo method has proven to be a robust technique to extract gross properties
in the energy region below the particle separation energy in terms of
nuclear level density (NLD) and $\gamma$-ray strength function ($\gamma$SF)~\cite{Schiller00,Lars11}.
The method is based on measuring the $\gamma$ decay after light ion reactions
where only one charged particle is ejected. In this way, one may collect $\gamma$ spectra tagged with
the excitation energy. Such coincidence data are then organized into an ($E_{\gamma}, E_x$) matrix, which
represents the basis for the Oslo method. Without going into details, this matrix is then corrected for
the detector response function~\cite{Gut96} and then utilized to extract the
first-generation (primary) $\gamma$ spectra~\cite{Gut87}. For illustration, the unfolded and primary
$\gamma$ matrices of $^{238}$Np are shown in Fig.~\ref{fig:matrix}

The primary $\gamma$-ray matrix, which represents the energy distribution
of the first emitted $\gamma$-ray transition in each cascade (see Fig.~\ref{fig:matrix} (b) ),
is assumed to be factorized by
\begin{equation}
P(E_{\gamma}, E_i) \propto   \rho(E_i-E_{\gamma}){\cal{T}}(E_{\gamma}) .\
\label{eqn:rhoT}
\end{equation}
The proportionality to $\rho(E_i -E_{\gamma})$ is in accordance with the Fermi's golden rule~\cite{dirac,fermi},
and the second factor, the $\gamma$-ray transmission coefficient ${\cal{T}}$, is assumed
to be independent of excitation energy according to the Brink hypothesis~\cite{brink}.
For dipole transitions, ${\cal{T}}$ relates to the $\gamma$SF by
\begin{equation}
f (E_{\gamma}) =\frac{1}{2 \pi}\frac{{\cal {T}}(E_{\gamma})}{ E_{\gamma}^3 }.
\end{equation}

\begin{figure}[t]
\begin{center}
\includegraphics[clip,width=0.9\columnwidth]{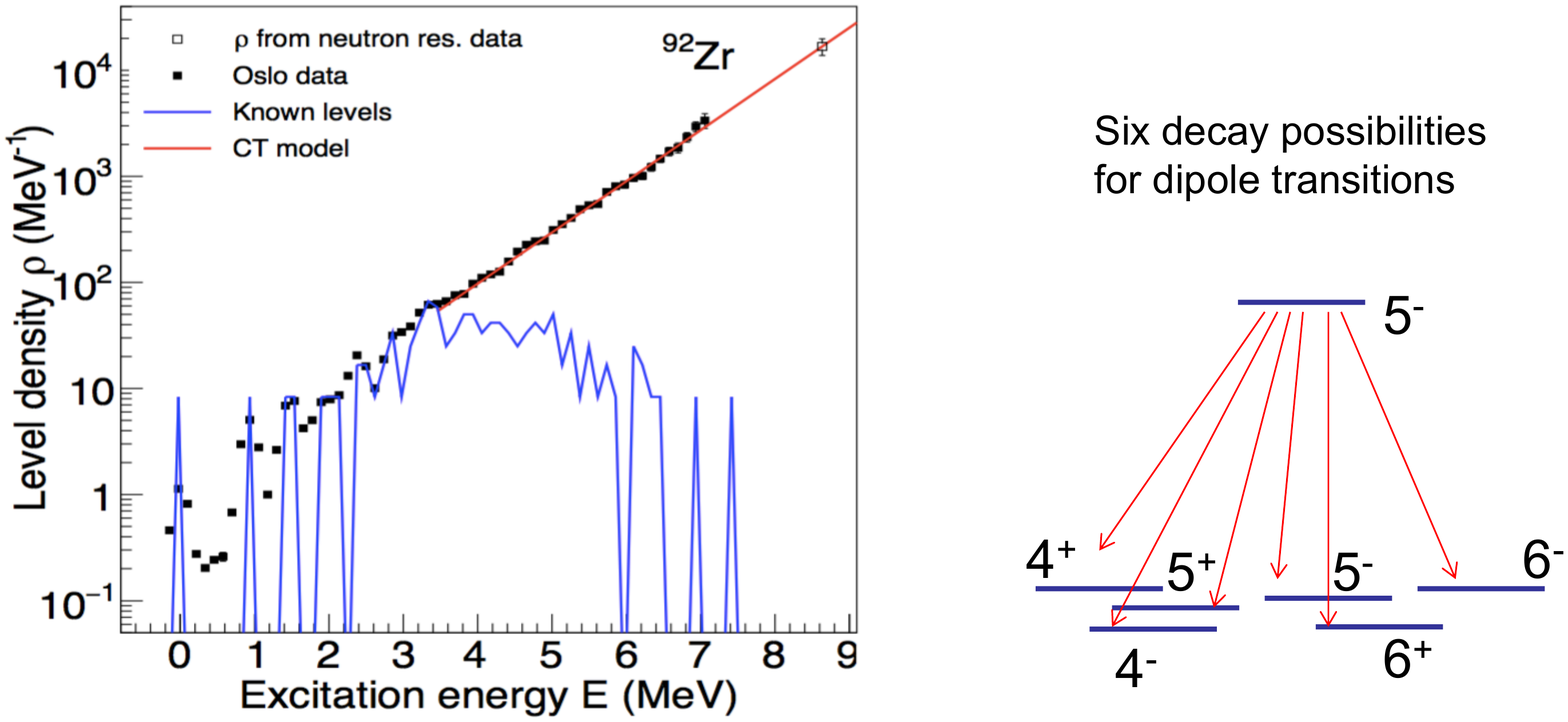}
\caption{(Color online) The level density extracted from the
$^{92}$Zr($p,p'$)$^{92}$Zr reaction based on the primary $P(E_{\gamma}, E_i)$ matrix.
In general, the dipole transition in the quasi-continuum
goes with six transitions: two stretched $M1$, two stretched $E1$,
one un-stretched $M1$ and one un-stretched $E1$ transition.}
\label{fig:six}
\end{center}
\end{figure}

In the standard Oslo method, we utilize the $\gamma$ spectra above a certain initial
excitation region (in the $^{238}$Np case, $E_{\rm min} = 3.03$ MeV) where the decay
is assumed to be of statistical nature. The upper excitation region is limited to
the neutron separation energy by $E_{\rm max} \approx S_n$.

By accepting the level density obtained with the fit of $\rho$ and ${\cal{T}}$ to
a large part of the $P$ matrix (the standard Oslo method),
we may investigate the transmission coefficient in more detail,
and thus the validity of the Brink hypothesis. We adopt
the solutions ${\cal T}$ and $\rho$ from Eq.~(\ref{eqn:rhoT}) and write
\begin{equation}
{\cal T} ( E_{\gamma}, E_i) \approx {\cal N}(E_i)
\frac{P(E_{\gamma}, E_i)}{\rho(E_i - E_{\gamma})}.
\label{eqn:neieg}
\end{equation}
with a normalization function given by
\begin{equation}
{\cal N}(E_i)=\frac{\int_0^{E_i} {\mathrm{d}} E_{\gamma } \, {\cal T}(E_{\gamma}) \rho(E_i-E_{\gamma})
}{\int_0^{E_i}{\mathrm{d}} E_{\gamma} \, P(E_{\gamma}, E_i)}.
\label{eqn:nei}
\end{equation}
Since ${\cal N}$ only depends on $E_i$, we may now study how
${\cal{T}}$ (and thus $f$) varies as function of $E_{\gamma}$ for
each excitation bin. An expression for ${\cal T} (E_{\gamma}, E_f)$
can be constructed in the same way as above~\cite{guttormsen2016}.

The fluctuations expected for the $\gamma$SF
are assumed to follow the $\chi ^2_{\nu}$ distribution where $\nu$ is the number of
 transitions $n$ included in the averaging for a certain energy $E_{\gamma}$.
The relative fluctuations of the $\chi ^2_{\nu}$ distribution
are given by $r= \sqrt{2/\nu}$. With the experimental information
on the level density $\rho$ (see e.g.~Fig.~\ref{fig:six}), we may count the
number of transitions expected from an initial to a final excitation energy bin.
For the standard Oslo method, a large part of the primary $\gamma$
matrix is utilized to obtain a good averaging. Here, all the rows
from $E_{\rm min}$ to $E_{\rm max}$ of the $P$ matrix are included:
\begin{equation}
n(E_{\gamma})= \Delta E ^2 \sum_{E_i = E_{\rm min}}^{E_{\rm max}}\sum_{I \pi} \sum_{L=-1}^{1}\sum_{\pi '}
\rho(E_i,I,\pi)\cdot \rho(E_i - E_{\gamma},I+L,\pi '),
\label{eqn:n}
\end{equation}
where $\Delta E$ is the energy-bin width for initial and final excitation energies.
If we assume the decay from only one initial excitation energy bin,
the number of transitions is given by
\begin{equation}
n(E_{\gamma},E_i)= \Delta E ^2 \sum_{I \pi} \sum_{L=-1}^{1}\sum_{\pi '}
\rho(E_i,I,\pi)\cdot \rho(E_i - E_{\gamma},I+L,\pi ').
\label{eqn:ni}
\end{equation}
A similar expression can be constructed for the cases when the decay
ends at a certain final excitation energy $E_f = E_i - E_{\gamma}$. The two formulae above have
to be taken with care if not all spins and parities are available. In
particular, for the direct dipole decay to the $0^+$ ground state,
only one transition appears, either one stretched $M1$ or $E1$ transition.
As illustrated in Fig.~\ref{fig:six}, this is contrary to six transitions in the high level
density region.

\begin{figure}[t]
\begin{center}
\includegraphics[clip,width=\columnwidth]{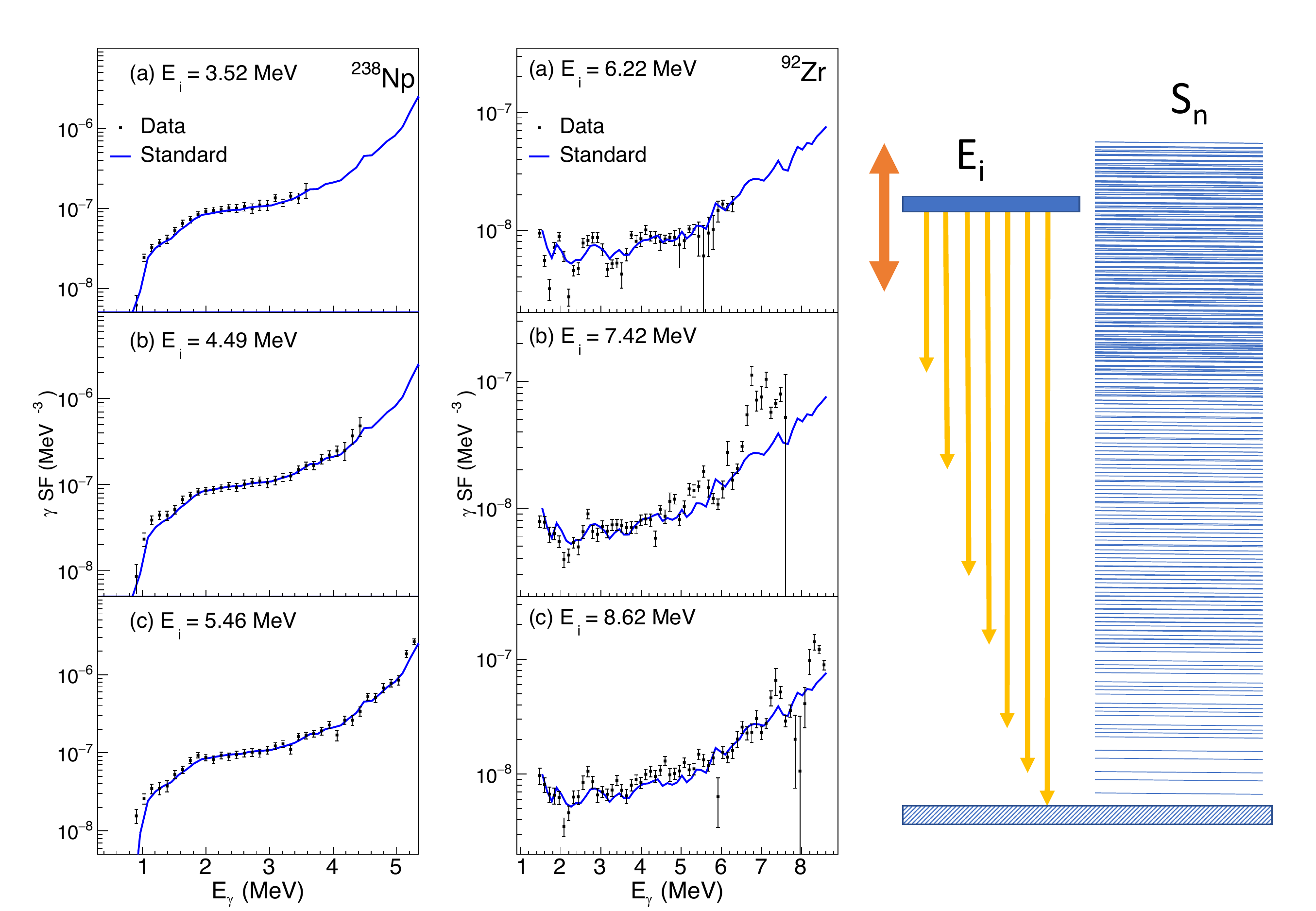}
\caption{(Color online) Gamma-ray strength functions of $^{238}$Np (left)
and $^{92}$Zr (right) extracted from various
initial excitation bins $E_i$ (see illustration). The data points with error bars
are from statistically independent data sets. The blue
lines are obtained by the standard Oslo method where a large part
of the primary $\gamma$-ray matrix has been exploited giving a better
averaging of the $\gamma$SF by including more transitions. We
observe significant fluctuations in the $^{92}$Zr case.}
\label{fig:rsfi}
\end{center}
\end{figure}

\section{Discussion}

In the following we will discuss the $\gamma$ decay in the quasi-continuum
of $^{238}$Np and $^{92}$Zr. Both
nuclei have been measured with high statistics allowing us to draw
conclusions on the dependence of the $\gamma$SFs with excitation energy.
The odd-odd $^{238}$Np nucleus has an extremely high level density of
$\approx 43$ million levels per MeV at $S_n = 5.488$ MeV. Also
at low excitation energy of $\approx 200$ keV we find $\approx 200$ levels per MeV.
Thus, for this ultimate system, we do not expect significant Porter-Thomas fluctuations
and the gBA hypothesis can be tested. However, the $^{92}$Zr nucleus
represents a system with $\approx 10000$ times less levels than $^{238}$Np
making it more difficult to draw conclusions on the validity of the
gBA hypothesis.

Figure \ref{fig:rsfi} shows $\gamma$SFs from three initial excitation bins
by use of Eq.~(\ref{eqn:neieg}). The blue curve is based on the standard
Oslo method of Eq.~(\ref{eqn:rhoT}) including many more primary $\gamma$ transitions.
In the case of $^{238}$Np, the individual $\gamma$SFs indeed
follow the same function obtained with the standard Oslo method.
The close resemblance with a constant excitation-independent
$\gamma$SF for these three initial excitations
is found for all the 22 excitation bins
between 3.03 and 5.46 MeV (not shown).

As expected, the $^{92}$Zr nucleus reveals strong fluctuations where
the individual $\gamma$SFs deviate more from the standard Oslo
method predictions (blue curve) than given by the statistical uncertainties.
Since this nucleus has lower NLD, we even find for certain $E_{\gamma}$-values
that none or very few $\gamma$ transitions exist within the
given initial excitation gate $E_i$. As an example, in the right lower panel (c) of Figure \ref{fig:rsfi}
a drop is seen in the $\gamma$SFs for $E_{\gamma} \approx 8.0$ MeV since no level exists
in $^{92}$Zr at a final excitation energy of
$E_f= E_i -E_{\gamma} = 8.62 - 8.00$ MeV = 0.62 MeV.

\begin{figure}[t]
\begin{center}
\includegraphics[clip,width=\columnwidth]{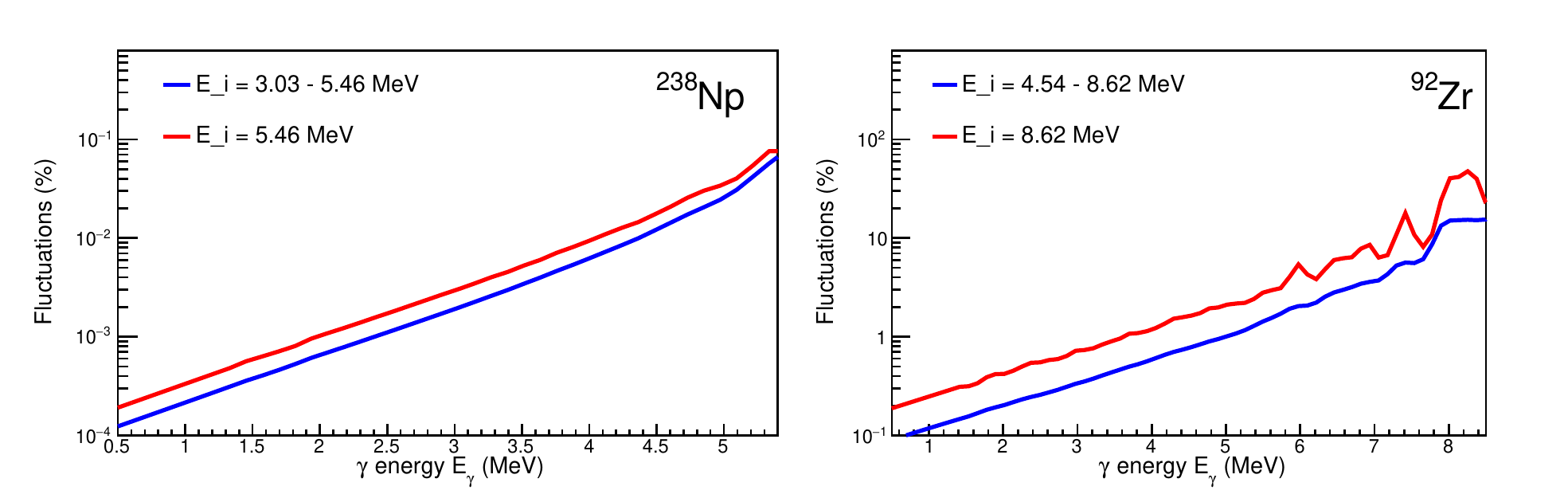}
\caption{(Color online) Relative fluctuations $r(E_{\gamma})$ in the $\gamma$ strength of $^{238}$Np (left)
and $^{92}$Zr (right) as function of $E_{\gamma}$.
The red lines are evaluated for a narrow initial high-energy
excitation bin, and the blue curves are the results based
on a broad initial excitation region (standard Oslo method).}
\label{fig:fluct}
\end{center}
\end{figure}

The findings of $^{92}$Zr versus $^{238}$Np raise a need
for a more quantitative prediction of the Porter-Thomas fluctuations.
If there is an insufficient averaging of the number of $\gamma$ transitions,
the fluctuations could be large and the individual $\gamma$SFs should
then be different. However, this may not invalidate the gBA hypothesis,
which predicts an underlying directive for the $\gamma$-decay probabillity.

In order to estimate the Porter-Thomas fluctuations, we apply the expressions
for the number of transitions described for the standard Oslo
method in Eq.~(\ref{eqn:n}) and for a specific initial excitation energy in
Eq.~(\ref{eqn:ni}). The relative fluctuations $r(E_{\gamma})$ of Fig.~\ref{fig:fluct} are
seen to increase exponentially with $E_{\gamma}$ for both nuclei.
We also note that the relative fluctuations are 2 - 3 times less for
the standard Oslo method compared to the fluctuations using an individual narrow initial
excitation bin. The fluctuations in $^{238}$Np are extremely low
going from $r\approx 0.0001$\% at low $E_{\gamma}$ and
up to  $\approx 0.1$\% at $S_n$. The $^{92}$Zr nucleus shows
higher fluctuations than the statistical errors, and up to
$r \approx 40$ \% at $E_{\gamma} \approx 8$ MeV. The individual
$\gamma$SFs of $^{92}$Zr (only 3 out of 34 shown in Fig.~\ref{fig:rsfi})
fluctuate around the $\gamma$SFs obtained with the standard Oslo method,
and thus support the validity of the gBA hypothesis also for this light
nucleus.

\section{Conclusion}

We have studied the $\gamma$SF between
excitation energy bins in $^{238}$Np and $^{92}$Zr. The $\gamma$ decay
in $^{238}$Np is independent on the excitation energy, and
thus validates the gBA hypothesis. In this case, the hypothesis
works not only for specific collective resonances, but for
the whole dipole $\gamma$SF below the neutron separation energy.

A necessity for testing the validity of the gBA hypothesis is that
the Porter-Thomas fluctuations are smaller than the experimental errors.
This means that the sampling of the $\gamma$SF should
include sufficiently many $\gamma$ transitions.
A technique has been developed using the measured level density
to estimate the number of transitions with a certain $\gamma$ energy.

For the case of $^{92}$Zr, large fluctuations are found that apparently
contradict the occurence of one unique $\gamma$SF for all excitation energies.
A closer examination reveals that these large deviations appear due to
few or non-existing levels for a given
$E_{\gamma}$ at certain excitation bins. In particular, huge
fluctuations appear for $\gamma$SFs including low-lying states,
e.g. the $I^{\pi}=0^+$ ground state. The individual $\gamma$SFs
seem to fluctuate around a common $\gamma$SF, which gives support
to the gBA hypothesis also for lighter nuclei.
\\
\\
{\bf Acknowledgements}
\\We would like to thank E.~A.~Olsen,
J.~C.~M{\"u}ller, A. Semchenkov, and J.~C. Wikne for providing high quality
experimental conditions. A.~C.~L. gratefully acknowledges funding
through ERC-STG-2014 under grant agreement no. 637686. G.~M.~T. gratefully
acknowledges funding of this research from the Research Council of Norway,
Project Grant No. 222287.


\begin{thebibliography}{99}

\bibitem{brink} D.M.~Brink, Doctoral thesis, Oxford University, 1955.
\bibitem{oslo2009} D.M.~Brink 2009, available online at {\em http://tid.uio.no/workshop09/talks/Brink.pdf}
\bibitem{axel} P.~Axel, Phys. Rev. {\bf 126}, 671 (1962).
\bibitem{Bartholomew} G.A. Bartholomew, E.D. Earle, A.J. Ferguson, J.W. Knowles, and M.A. Lone, Adv. Nucl. Phys. {\bf 7}, 229 (1973).
\bibitem{stefanon1977} M. Stefanon and F. Corvi, Nucl.~Phys.~A {\bf 281}, 240 (1977).
\bibitem{raman1981} S. Raman, O. Shahal, and G.G.~Slaughter, Phys.~Rev.~C {\bf 23}, 2794 (1981).
\bibitem{kahane1984} S. Kahane, S. Raman, G.G.~Slaughter, C.Coceva and M. Stefanon, Phys.~Rev.~C {\bf 30}, 807 (1984).
\bibitem{islam1991} M.A. Islam, T.J. Kennett, and W.V. Prestwich, Phys.~Rev.~C {\bf 43}, 1086 (1990).
\bibitem{kopecky1990} J.~Kopecky and M.~Uhl, Phys.\ Rev.\ C \bf 41\rm, 1941 (1990).
\bibitem{netterdon2015} L. Netterdon, A. Endres, S. Goriely, J. Mayer, P. Scholz, M. Spieker, and A. Zilges, Phys. Lett. B {\bf 744}, 358 (2015).
\bibitem{johnson2015} C.W. Johnson, Phys. Lett. B {\bf 750}, 72 (2015).
\bibitem{misch2014} G. Wendell Misch, George M. Fuller, and B. Alex Brown, Phys.~Rev.~C {\bf 90}, 065808 (2014).
\bibitem{koeling1978} T. Koeling, Nucl.~Phys.~A {\bf 307}, 139 (1978).
\bibitem{horing1992} A. H\"{o}ring and H.A. Weidenm\"{u}ller, Phys.~Rev.~C {\bf 46}, 2476 (1992).
\bibitem{gu2001} J.Z. Gu, H.A. Weidenm\"{u}ller, Nucl.~Phys.~A {\bf 690}, 382 (2001).
\bibitem{betak2001} E. B\u{e}t\'{a}k, F. Cvelbar, A. Likar, and T. Vidmar, Nucl.~Phys.~A {\bf 686}, 204 (2001).
\bibitem{hussein2004} M.S. Hussein, B.V. Carlson and L.F. Canto, Nucl.~Phys.~A {\bf 731}, 163 (2004).
\bibitem{PT} C.E. Porter and R.G. Thomas, Phys. Rev. {\bf 104}, 483 (1956).
\bibitem{Schiller00} A.~Schiller, L. Bergholt, M.~Guttormsen, E. Melby, J.~Rekstad, and S.~Siem, Nucl. Instrum. Methods Phys. Res. A {\bf 447}, 494 (2000).
\bibitem{Lars11} A.C.~Larsen {\em et al.}, Phys.\ Rev.\ C \bf 83\rm, 034315 (2011).
\bibitem{Gut96} M.~Guttormsen, T.~S.~Tveter, L.~Bergholt, F.~Ingebretsen, and J.~Rekstad, Nucl.\ Instrum.\ Methods Phys.\ Res.\ A \bf 374\rm, 371 (1996).
\bibitem{Gut87} M.~Guttormsen, T.~Rams{\o}y, and J.~Rekstad, Nucl.\ Instrum.\ Methods Phys.\ Res.\ A \bf 255\rm, 518 (1987).
\bibitem{238Np} T.G. Tornyi {\em et al.}, Phys.\ Rev.\ C \bf 89\rm, 0443232 (2014).
\bibitem{guttormsen2016} M. Guttormsen, A. C. Larsen, A.~G{\"o}rgen, T.~Renstr{\o}m, S. Siem, T. G. Tornyi and G. M. Tveten, Phys.~Rev.~Lett. {\bf 116}, 012502 (2016).
\bibitem{dirac} P.A.M. Dirac, "The Quantum Theory of Emission and Absortion of Radiation". Proc. R. Soc. Lond. A 1927 114, 243-265.
\bibitem{fermi} E. Fermi, Nuclear Physics. University of Chicago Press (1950).


\end{thebibliography}
\end{document}